\documentclass[twocolumn,aps,pra,epsfig,graphics,showpacs]{revtex4}
\usepackage{graphicx}
\usepackage{amsmath}
\usepackage{amsfonts}
\usepackage{amssymb}

\newcommand{\di}{\partial}

\begin{document}

\title{Generation of Squeezed States of Nanomechanical Resonators by
Reservoir Engineering}
\author{P. Rabl$^{1}$, A. Shnirman$^{2}$ and P. Zoller$^{1}$}
\affiliation{$^{1}$ Institute for Theoretical Physics, University of Innsbruck, and\\
Institute for Quantum Optics and Quantum Information of the Austrian Academy
of Science, 6020 Innsbruck, Austria\\
$^{2}$ Institut f{\"u}r Theoretische Festk{\"o}rperphysik , University of
Karlsruhe, Germany }

\begin{abstract}
An experimental demonstration of a non-classical state of a nanomechanical resonator is still an outstanding task. In this paper we show how
the resonator can be cooled and driven into a squeezed state by a bichromatic microwave coupling to a charge qubit. The stationary oscillator state 
exhibits a reduced noise in one of the quadrature components by a factor of 0.5 - 0.2. These values are obtained for a 100 MHz resonator
with a Q-value of 10$^4$ to 10$^5$ and for support temperatures of T $\approx$ 25 mK. We show that the coupling to the charge qubit can also be used to 
detect the squeezed state via measurements of the excited state population. Furthermore, by extending this measurement procedure 
a complete quantum state tomography of the resonator state can be performed. This provides a universal tool to detect a large variety of different
states and to prove the quantum nature of a nanomechanical oscillator.  
\end{abstract}

\maketitle

\section{Introduction}

With fabrication of nanomechanical resonators with fundamental frequencies
from 100 MHz up to 1 GHz~\cite{ClelandBook,ClelandRoukes,RoukesNature} the
demonstration of their quantum nature, in
particular, the creation of non-classical states of these mesoscopic
systems, have attracted a lot of interest~\cite{Wybourne,Blencowe_S,Schwab_Entangle,Schwab_Measure,Eisert}.
Apart from the fundamental interest in the study of these systems,
nanomechanical resonators are also of great importance for technical
applications. While micron-sized cantilevers are already used to perform
atomic force measurements, their noise properties which set the limits for
the sensitivity of force detection \cite{Braginsky,MEMS}
can be improved by scaling them down to the nanometer regime. Residual
thermal fluctuations can then be reduced to the quantum limit by active cooling schemes as proposed in~%
\cite{Schwab_Feedback, Ignacio, Martin_Cool}. 

A further reduction of the quantum fluctuations below the standard quantum limit can be achieved by 
squeezing the resonator mode. The idea to use squeezed states for measurements beyond the standard
quantum limit~\cite{Caves} appeared first in the context of the detection of gravitation waves.
In principle it applies to any system where a weak classical force which has to be measured 
acts on a quantum-mechanical oscillator. The measurement consists of three stages. In the first 
stage one prepares the oscillator in a squeezed state, so that the dispersion of one of its quadratures
is reduced below the quantum limit. Next, one allows the measured force to act on the oscillator. 
At last one measures the squeezed quadrature, without touching the other one. To apply 
the squeezing ideas to the nanomechanical systems, all three stages have to be implemented.
In particular, for the third stage, one has to design a setup in which only one of the quadratures is being 
measured. In this paper we restrict ourselves to the first stage, i.e., the preparation 
of the squeezed state of a nanomechanical oscillator. We show that by coupling the oscillator to a 
Cooper Pair Box (Josephson qubit) and irradiating the system by bichromatic, phase coherent 
microwaves one can ``cool'' the oscillator down to the squeezed state. 

We start by reminding the reader about the basics of squeezing and discuss two different 
ways to achieve it. Then we analyze the coupled CPB - oscillator system and show 
that the cooling into the squeezed state is feasible. At last we discuss possible ways 
to detect the squeezed state.

\section{Squeezed States and Reservoir Engineering}

\begin{figure}[tbp]
\begin{center}
\includegraphics[width=0.25\textwidth]{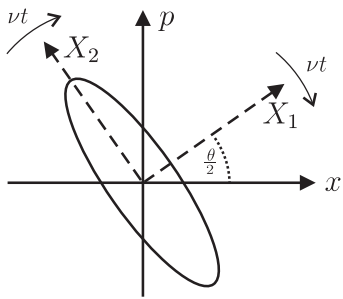}
\end{center}
\caption{Representation of the squeezed state $|\,\protect\epsilon %
\!=\! r e^{i\protect\theta}\rangle$ in phase space spanned by the dimensionless position and
momentum coordinates. The error ellipse indicates the reduced/enhanced fluctuations for the quadrature components
$X_1$/$X_2$.}
\label{pic:ErrorEllipse}
\end{figure}

For a harmonic oscillator with a Hamiltonian $H= \hbar \nu \, a^\dag a$, where $%
a$ and $a^\dag$ are the usual creation and annihilation operators, a general
class of Gaussian minimum-uncertainty squeezed states is defined by~\cite{Walls}
\begin{equation}\label{eq:Definition}
|\alpha,\epsilon\rangle=\hat D (\alpha)\hat S(\epsilon)|0\rangle
\end{equation}
Here $\hat D(\alpha)\!=\!\mathrm{exp}(\alpha a^\dag-\alpha^* a)$ is a
displacement operator, and $\hat S(\epsilon)\!=\!\mathrm{exp}(\frac{%
\epsilon^*}{2} a^2 -\frac{\epsilon}{2} a^{\dag 2})$ denotes the squeezing
operator. We refer to the state $|\epsilon\rangle\equiv
|\alpha\!=\!0,\epsilon\rangle$ as the squeezed vacuum state. The absolute value
of the complex number $\epsilon\!=\!r e^{i\theta}$ is called the squeezing
parameter. For these squeezed states the quadrature components, $X_{1,2}$,
defined by $a=(X_1+iX_2)e^{i\theta/2}$ fulfill the uncertainty relation $%
\Delta X_1 \Delta X_2 \geq\frac{1}{4}$, where the variance of one component,
$\Delta X_1=e^{-r}/2$ is reduced below the standard quantum limit of 1/2,
whereas the noise in the other component is enhanced, $\Delta X_2=e^{+r}/2$.
This property can be exploited to improve the sensitivity of measurements.
It is important to note that this asymmetric distribution of the noise is
stationary only in a frame rotating with the frequency $\nu$ of the
harmonic oscillator, since an initial squeezed state $|\epsilon\rangle$
evolves in time as $|\epsilon e^{-i2\nu t}\rangle$. Therefore, the error
ellipse rotates in phase space with the oscillator frequency, $\nu$, as
shown in Fig.~\ref{pic:ErrorEllipse}.

There are several ways to generate a squeezed state of a harmonic
oscillator. A familiar method in quantum optics~\cite{SqueezedLight}
%\textbf{PZNote:} \emph{When
%we refer to quantum optics, we need somewhere reference to the exp by
%Kimble, Polzik, and Luis Oroszco. This should also be cited in the
%introduction.}
is to use a parametrically driven non-linear potential corresponding to the
Hamiltonian
\begin{equation}
H= \hbar \nu a^\dag a - i\hbar\lambda(a^{\dag 2}e^{-i\omega_p t}- a^2
e^{i\omega_p t})\,.
\end{equation}
By going to a rotating frame, i.e.~transforming the time-dependence away and
assuming a parametric pump field frequency $\omega_p=2\nu$, the Hamiltonian
is simply given by $H_{\mathrm{I}}=i\hbar\lambda ( a^2-a^{\dag 2})$. Starting
from the ground state of the harmonic oscillator, $|0\rangle_{\mathrm{I}}$
the time evolution with $H_{\mathrm{I}}$ produces the squeezed state $%
|\epsilon\!=\!2\lambda t\rangle_{\mathrm{I}}$. The application of this
method for mechanical resonators has been proposed in Ref.~\cite{Blencowe_S}%
, but the requirements, a sufficiently strong nonlinearity to overcome the
losses, and an initial state close to the ground state, are not easily met.

A second method, which we will elaborate on below, is to ``engineer'' an
appropriate coupling to the environment such that a dissipative dynamics
drives the harmonic oscillator into a squeezed state, i.e.~ we ``cool'' the
oscillator mode to a squeezed state. This \emph{reservoir engineering} has
been first proposed in Ref.~\cite{Zoller_Squeeze,Zoller_Eng} in the context
of ion traps, and has been experimentally implemented in part by the Ion
Trap Group at NIST in Boulder~\cite{Wineland}. This reservoir engineering can be achieved, for example, by
coupling the oscillator to a dissipative two level system (TLS), where the
form of the coupling determines the stationary state.

The simplest (although trivial) example is provided by a reservoir which
cools the oscillator to the ground state $|0\rangle$. We assume that the
oscillator is coupled to a two-level system with ground and excited state $%
|g\rangle$, $|e\rangle$ according to the Hamiltonian $H_{\mathrm{I}%
}=g(a\sigma_+ + a^\dag \sigma_-)$, where we use Pauli spin notation $%
\sigma_+ \equiv |e\rangle \langle g|$ \emph{etc.}. We furthermore assume
that the two-level system decays from the excited state to the ground state
with a rate $\Gamma$. The time evolution in the interaction picture is then
described by the master equation
\begin{equation}  \label{eq:IdealMaster}
\frac{d\rho}{dt}=-i[H_{\mathrm{I}},\rho]+\frac{\Gamma}{2}\left(2\sigma_{-}%
\rho\sigma_{+}-\sigma_{+}\sigma_{-}\rho -\rho\sigma_{+}\sigma_{-}\right) \; .
\end{equation}
For long times the system thus evolves to the steady state $%
\rho=|0\rangle\langle 0 |\otimes |g\rangle\langle g|$ since $%
|0\rangle|g\rangle$ is a ``dark state'' of the system Hamiltonian, i.e.~$H_{%
\mathrm{I}}|0\rangle|g\rangle=0$.

\begin{figure}[tbp]
\begin{center}
\includegraphics[width=0.45\textwidth]{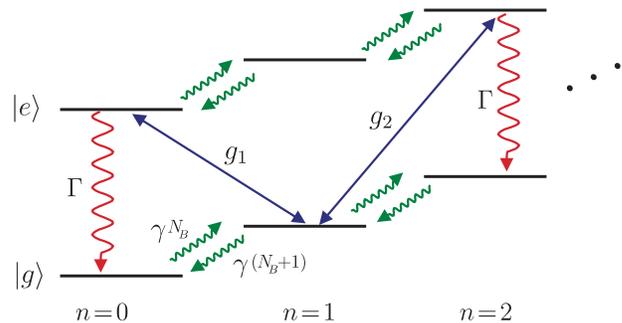}
\end{center}
\caption{Coherent and incoherent processes as described by master equation~
\eqref{eq:Master}. The excitations of the qubit coherently decrease ($g_1$,``red sideband'')
and increase ($g_2$,``blue sideband'') the phonon
number of the oscillator. The energy of the excite state is then dissipated with the decay rate of the TLS, $\Gamma$.
These two processes relax the system into the desired state, $\rho=|\epsilon,g\rangle\langle \epsilon,g|$ while the coupling to the thermal phonon bath, $\gamma$
thermalizes the oscillator state.}
\label{pic:LevelScheme}
\end{figure}

For a general Hamiltonian of the form $H_{\mathrm{I}}=\hat F \sigma_+ +\hat
F^\dag \sigma_-$ where $\hat F$ is a function of $a$ and $a^\dag$ only, the
dynamics of master equation~\eqref{eq:IdealMaster} ``cools'' the system into
the state $\rho=|\psi\rangle\langle \psi |\otimes |g\rangle\langle g|$. The
stationary oscillator state $|\psi\rangle$ is determined by $\hat F
|\psi\rangle =0$. A squeezed vacuum state, $|\epsilon\!=\!re^{i%
\theta}\rangle$ obeys the relation
\begin{equation}
\left(a \cosh(r)+ a^\dag \sinh(r)e^{-i\theta}\right)|\epsilon\rangle=0\,,
\end{equation}
and thus we choose $\hat F$ to be of the form ~\cite{Zoller_Squeeze}
\begin{equation}  \label{eq:IdealSqueezedState}
\hat F=g_1 a +g_2 e^{-i\theta}a^\dag\, ,
\end{equation}
where $g_1$ and $g_2$ are related by $r\!=\!\mathrm{atanh}(g_2/g_1)$.
For a single trapped ion driven by laser light and decaying via spontaneous emission such
a Hamiltonian, $H_{\rm I}$ with $\hat F$ given in Eq.~(\ref{eq:IdealSqueezedState}) can be
constructed by applying two laser beams, one detuned to the ``red sideband''
($\omega=\omega_{ge}-\nu$) and a weaker one detuned to the ``blue sideband''
($\omega=\omega_{ge}+\nu$) of the two-level transition frequency $%
\omega_{ge} $. This is illustrated in in Fig.~\ref{pic:LevelScheme}. For a detailed
explanation the reader is referred to the review article by Leibfried \textit{et al}.~\cite{IonReview}.

In the following section we show how such a Hamiltonian can be realized for
a nanomechanical resonator coupled to a Cooper pair box, which plays the
role of the dissipative two-level system ~\cite{Martin_Cool}.

For a mechanical resonator the coupling of the oscillator mode
to the finite temperature phonon bath of the support has to be taken into
account, leading to additional contributions in the master equation. As we discuss below, 
in the limit of high resonator frequencies where the rotating wave approximation (RWA) is valid
the master equation is 
\begin{equation}  \label{eq:Master}
\begin{split}
\frac{d\rho}{dt}=-&i[H_{\rm I},\rho]+\frac{\Gamma}{2}\left(2\sigma_{-}\rho%
\sigma_{+}-\sigma_{+}\sigma_{-}\rho -\rho\sigma_{+}\sigma_{-}\right) \\
&+(N_{\mathrm{B}}+1)\frac{\gamma}{2}\left(2a\rho a^\dag-a^\dag a \rho - \rho
a^\dag a \right) \\
&+N_{\mathrm{B}}\frac{\gamma}{2}\left(2a^\dag\rho a- a a^\dag \rho - \rho a
a^\dag \right) \, .
\end{split}%
\end{equation}
The new terms in the master equation describe the heating and dissipation of
the resonator with rate $\gamma$, which relaxes to a thermal state with mean
phonon occupation $\langle a^\dagger a \rangle = N_{\mathrm{B}}$. This will
degrade the squeezing.

\section{The Model}

\begin{figure}[b]
\begin{center}
\includegraphics[width=0.45\textwidth]{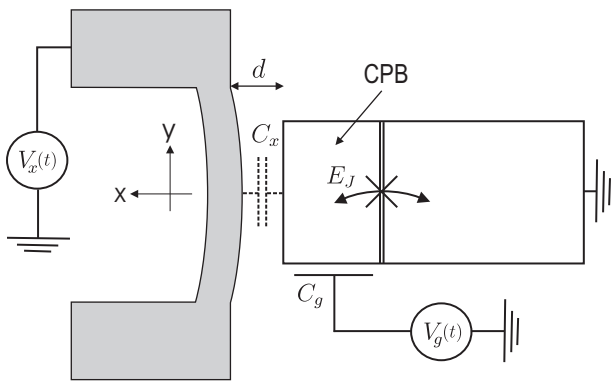}
\end{center}
\caption{Setup: The mechanical resonator (thin, vertical bar) is biased by the voltage $%
V_x$ to achieve a coupling to the Cooper pair box (CPB) via the capacitance $%
C_x$. The state of the CPB can be controlled independently by the gate
voltage, $V_g$ and the Josephson energy, $E_J$. }
\label{pic:Setup}
\end{figure}
We consider a nanomechanical resonator which is placed close to a Cooper
pair box (CPB) as shown in Fig.~\ref{pic:Setup}. Similar systems have been
considered in the context of cooling~\cite{Martin_Cool} and the generation
of entangled states~\cite{Schwab_Entangle}. The energy spectrum of the CPB
alone is controlled by the gate voltage, $V_g$ and the Josephson energy, $E_J$~\cite{ShnirmanRev}.
To obtain a coupling between the two systems the voltage $V_{x}$ is applied on
the resonator, leading to a position dependent interaction via the
capacitance $C_x(x)$. The whole system is described by the Hamiltonian
\begin{equation}\label{eq:FullHamiltonian}
H=\frac{(Q-Q_{g})^2}{2C_{\Sigma}}-E_{J}\cos(\phi) + \hbar \nu a^{\dag}a\, .
\end{equation}
Here $Q_{g}=C_{g}V_{g}+C_{x}V_{x}$ is the total gate charge, $C_{\Sigma}$ is
the total capacitance of the island and $\nu$ is the frequency of the
fundamental flexural mode of the resonator. We decompose the voltages into a
sum of a constant and a time dependent part, $V_i=V_i^{0}+V_i(t)$ and
therefore $Q_{g}=2e(n_{g}^{0}+n_{g}(t))$. For the desired interaction and to minimize
relaxation (see below) the system is operated
close to the degeneracy point, $n^{0}_{g}=n+1/2$, $n_{g}(t)<<1$ and
we reduce the CPB to an effective two level system with the basis states
\begin{equation}
\begin{split}
|g\rangle&=(|n\rangle+|n+1\rangle)/\sqrt{2}\,, \\
|e\rangle&=(|n\rangle-|n+1\rangle)/\sqrt{2}\,.
\end{split}%
\end{equation}
The Hamiltonian~\eqref{eq:FullHamiltonian} simplifies to
\begin{equation}  \label{eq:Hamiltonian1}
\begin{split}
H=&+\frac{E_{J}}{2}\sigma_{z}+\,4E_{c}(x)n_{g}(t,x)\sigma_{x} \\
&+ E_{c}(x)(1+4n_{g}(t,x)^2) + \hbar \nu a^\dag a \, .
\end{split}%
\end{equation}
Note that $E_{c}(x)=e^2/2C_{\Sigma}$ as well as $n_{g}(t,x)$ depend on the
resonator coordinate x via the capacitance $C_{x}(x)$. If we expand $C_{x}$
up to the first order in x/d, where d is the distance between the resonator
and the CPB we obtain
\begin{eqnarray*}
E_{c}(x)&=& E_{c}+E_{c}\frac{C_{x}^{0}}{C_{\Sigma}}\frac{x}{d}+\mathcal{O}(%
\frac{x^2}{d^2})\, , \\
n_{g}(t,x)&=&\frac{1}{2e}\left( C_{g}V_{g}(t)+C_{x}^{0}V_{x}(t)(1-\frac{x}{d}%
)\right)+\mathcal{O}(\frac{x^2}{d^2}) \, .
\end{eqnarray*}
Re-substituting these expansions into equation \eqref{eq:Hamiltonian1} and
absorbing small shifts of the equilibrium position of the resonator in a redefinition
of $a$ and $a^\dag$ we obtain the following contributions.
\begin{equation}  \label{eq:ExpandHamiltonian}
\begin{split}
H =&\,+\frac{E_J}{2}\sigma_z+ \frac{4E_{c}}{2e}%
(C_{g}V_{g}(t)+C_{x}^{0}V_{x}(t))\sigma_{x} \\
&+\frac{4E_{c}}{2e}\left(\frac{C_x^0}{C_{\Sigma}}
C_{g}V_{g}(t)-C_{x}^{0}V_{x}(t)(1-\frac{C_{x}^{0}}{C_{\Sigma}})\right)\frac{x%
}{d}\,\sigma_{x} \\
&+\hbar \nu a^\dag a + \mathcal{O}(\frac{x^2}{d^2})\,.
\end{split}%
\end{equation}
The second term in Eq.~%
\eqref{eq:ExpandHamiltonian} leads to direct excitations of the charge
qubit. To avoid these excitations we choose the voltage signals such that
they do not alter the gate charge, $Q_g(t)/2e=C_{g}V_{g}(t)+C_{x}^{0}V_{x}(t)\approx 0$. For the systems under
consideration the expansion parameter, $\langle x\rangle/d$ is in the order of $%
10^{-6}$ and we can neglect all higher order terms in H.

With these approximations the system Hamiltonian~\eqref{eq:ExpandHamiltonian} reduces to
\begin{equation}
H=+\frac{E_{J}}{2}\sigma_{z}-\lambda(t)(a+a^{\dag})\sigma_{x}+\hbar\nu a^\dag a \,.
\end{equation}
Up to now the result is valid for arbitrary driving signals. For the
generation of squeezed states we choose the driving voltage to be of the
form
\begin{equation}
V_x(t)= V_1\cos\left(\frac{(E_{J}-\nu)t}{\hbar}\right)+V_2\cos\left(\frac{%
(E_{J}+\nu)t}{\hbar}+\theta\right)\,.
\end{equation}
It consists of a part tuned to the red sideband and a part tuned to the blue
sideband of the qubit transition frequency related by a fixed phase difference, $%
\theta$. Finally we perform a transformation into the interaction picture with respect to $%
H_0=E_J/2 \,\sigma_z +\hbar \nu \,a^\dag a$. Under the assumption $|\lambda(t)|\ll \hbar\nu$ the RWA can be applied 
and we end up with
\begin{equation}  \label{eq:SqueezeHamiltonian}
H=\hbar\left(g_1 a+ g_2 e^{-i\theta} a^\dag\right)\sigma_{+}+\hbar\left(g_2e^{i\theta} a+ g_1
a^\dag\right)\sigma_{-} \,,
\end{equation}
with the parameters
\begin{equation}
g_i=-2E_C\frac{x_0}{d}\frac{C_{x}^{0}V_{i}}{2e}\, ,
\end{equation}
where $x_0=\sqrt{\hbar/2m\nu}$ is the extension of the resonator ground state.

\emph{Discussion.} For typical parameters values $C_{x}^{0}\approx
2\times10^{-17}$ F, $E_{C}\approx 40 $ GHz, $x_0/d\approx 10^{-6}$, we
obtain a value for the coupling strength of about $g_{i}\approx 5$ MHz for
driving voltages still below 1 V. This value is also consistent with the
approximations we have made ($n_{g}(t)\!\ll \!1$, RWA, \dots ), assuming a resonator with
a fundamental frequency $\nu\geq100$ MHz.

In practice, a perfect realization of the balance condition $Q_g(t)=0$ is impossible
which leads to direct excitations of the charge qubit. However, an accuracy
in the control of the voltages in the order of $x_0/d$ is sufficient to neglect
this term since the applied voltages are detuned from the qubit transition
frequency by $\nu$.

Insufficient precision in the knowledge of the oscillator frequency as well
as the qubit transition frequency leads to unavoidable detunings for the
applied driving fields. Their effect is taken into account by adding the terms $\delta_x a^\dag
a+\delta_{cq}|e\rangle\langle e|$ to Hamiltonian~%
\eqref{eq:SqueezeHamiltonian}. A detuning from the exact resonator
frequency, $\delta_x$ destroys perfect squeezing because the ideal state
$|\epsilon\rangle$ is not an eigenstate of $a^\dag a$. Since a measurement
of the resonator frequency with a resolution of a few ppm can be achieved~%
\cite{Schwab_Measure}, $\delta_x$ is less than 1 kHz. The residual imperfection is a small effect
compared to the influence of the finite Q-value and can therefore be neglected. The
detuning from the charge qubit transition frequency, $\delta_{cq}$ is less
crucial since it does not affect the steady state. For $\delta_{cq}<g_i$ it
only slightly changes the excitation probabilities of the qubit. A
measurement with the required precision has been reported by Vion \textit{et al}~\cite{Vion}.

\emph{Damping.} 
Apart from the unitary evolution given by H, the coupling to the environment
provides the dissipative part of the system dynamics.
While a finite decay rate of the charge qubit is crucial for \emph{reservoir engineering} the damping of the
resonator mode sets the limits of this method. Here we give a brief discussion of the dominant effects in our system
due to the influence of the environment.

The mechanisms of dissipation in superconducting qubits have not yet been fully investigated.
The early experiments~\cite{Nakamura_Nature} reported decoherence times of order several nano-seconds.
This has been attributed to the effect of the low frequency ($1/f$) noise. In Ref.~\cite{Vion} it was demonstrated
that this effect can be substantially reduced by operating at special symmetry points (degeneracy points). 
The decoherence time of $500$ ns was achieved. Moreover, it became clear that a substantial part
of the decoherence at such points is due to the energy relaxation ($T_1$ in NMR) processes. 
Here, for simplicity, we assume that only the energy relaxation is important at the symmetry point. 
It is provided by the high frequency modes of the environment.
One of the possible relaxation channels is the electro-magnetic environment in the external circuits
which created fluctuations of the gate voltages. Taking into account these fluctuations by substituting $
V_i\rightarrow V_i+\delta V_i$ we obtain the additional terms $\frac{4E_C}{2e%
}C_i\delta V_i \sigma_x$ due to this voltage noise. Assuming equal noise
characteristics for $\delta V_g$ and $\delta V_x$ this implies a decay rate
\begin{equation}
\Gamma_{e\rightarrow g}=\frac{e^2}{\hbar^2} \frac{C_x^2+C_g^2}{C_{\Sigma}^2}%
S_{V}(+E_J/\hbar).
\end{equation}
For an external impedance $Z(w)$ the noise spectrum of the voltage is given
by $S_V(\omega)=2\mathrm{Re} Z(w)\hbar \omega [1-e^{-\hbar \omega/k_B
T}]^{-1}$. Because the temperatures reached with dilution refrigerators are
in the order of $10-50$ mK which is much smaller than $E_J/k_B$ excitations of
the qubit can be neglected and the decay rate simplifies to
\begin{equation}
\Gamma=\Gamma_{e\rightarrow g}=\pi \frac{C_x^{2}+C_{g}^2}{C_\Sigma^2}\frac{R%
}{R_Q}\frac{E_J}{\hbar}\, ,
\end{equation}
where R$_Q=h/4e^2$ is the resistance quantum. The decay
rate can be adjusted by the gate capacitance, $C_g$ and has typically values
of 1-10 MHz~\cite{Vion}.

The dominant mechanism for the damping of the resonator mode is the
coupling to the phonon modes of the support. This leads to a finite decay
rate $\gamma =\nu /Q$, where Q is the quality factor of the resonator. In
contrast to the charge qubit the temperature of the environment is higher or comparable
to the oscillator frequency. Therefore, the phonon modes at the resonator frequency have a
non-zero occupation, $N_{B}=[e^{\hbar \nu /k_{B}T}-1]^{-1}$ and cause downward and upward transitions.
For temperatures of $10-50$ mK the
oscillator ($\nu $ $=100$ MHz) has an equilibrium occupation number $N_{B}\approx 2-10$.

Together with the Hamiltonian~\eqref{eq:SqueezeHamiltonian} the decay rates $\Gamma$, $\gamma$ and the bath occupation
number $N_B$ lead to a dissipative dynamics of the system described by the master equation~\eqref{eq:Master}.

\section{Results}

We are interested in the properties of the steady state solution of master
equation~\eqref{eq:Master}. For a characterization of the stationary state
we concentrate on the variance of the $X_1$ quadrature component, $(\Delta X_1)^2_{\rho_s}=\langle
X_1^2\rangle-\langle X_1 \rangle^2$ where the average is taken with respect to
the steady state density matrix, $\rho_s$. We compare the variance $\Delta X_1$ with the
zero point fluctuations, $(\Delta X_1)^2_0=\langle 0| X_1^2|0\rangle$ and define the ratio
\begin{equation}
\mathcal{R}=\frac{(\Delta X_1)_{\rho_s}}{(\Delta X_1)_0}
\end{equation}
 as a measure for the degree of squeezing.

\begin{figure}[tbp]
\begin{center}
\includegraphics[width=0.45\textwidth]{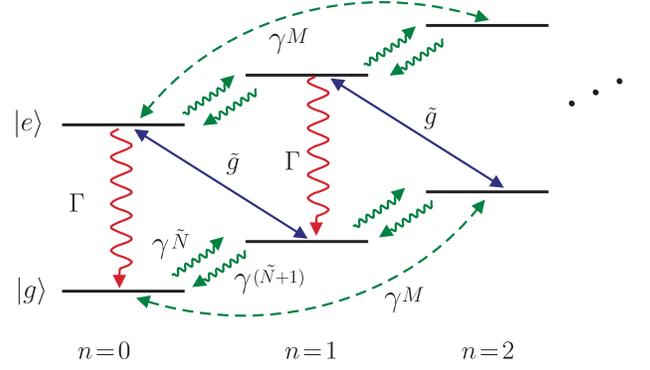}
\end{center}
\caption{Coherent and incoherent processes in the squeezed frame as
described by master equation~\eqref{eq:TransMaster}. In this picture the qubit is only excited
on the red sideband ($\tilde g$). Followed by a spontaneous decay ($\Gamma$) these transitions cool the oscillator to its ground state.
The cooling is compensated by the squeezed heat bath which apart from the enhanced heating rate ($\gamma \tilde N$)
also induces coherences between the states $|n\rangle$ and $|n\pm2\rangle$ ($\gamma M$).}
\label{pic:LevelScheme2}
\end{figure}

To study the effect of the different terms in Eq. \eqref{eq:Master} it is
convenient to look at the master equation in the squeezed frame, i.e. we
perform the unitary transformation, $U=\hat S(\epsilon)$ where the value of $%
\epsilon=re^{i\theta}$ is chosen according to Eq.~\eqref{eq:IdealSqueezedState}. For the
transformed density operator, $\tilde \rho=U^\dag \rho U$ we obtain the
equation
\begin{equation}  \label{eq:TransMaster}
\begin{split}
\frac{d\tilde\rho}{dt}=&-i[\,\tilde g(a\sigma_+
+a^\dag\sigma_-),\tilde\rho\,] \\
&+\frac{\Gamma}{2}\left(2\sigma_{-}\tilde\rho\sigma_{+}-\sigma_{+}\sigma_{-}%
\tilde\rho -\tilde\rho\sigma_{+}\sigma_{-}\right) \\
&+\frac{\gamma}{2}(\tilde N+1)\left(2a\tilde\rho a^\dag-a^\dag
a\tilde\rho-\tilde\rho a^\dag a \right) \\
&+\frac{\gamma}{2}\tilde N \left(2a^\dag\tilde\rho a- a a^\dag \tilde\rho -
\tilde\rho a a^\dag\right) \\
&-\frac{\gamma}{2}M\left(2a\tilde\rho a-a a\tilde\rho-\tilde\rho a a \right)
\\
&-\frac{\gamma}{2}M^*\left(2a^\dag\tilde\rho a^\dag-a^\dag
a^\dag\tilde\rho-\tilde\rho a^\dag a^\dag \right)
\end{split}%
\end{equation}
where $\tilde g=|g_1|/\cosh(r)$ and
\begin{eqnarray*}
\tilde N &=&(N_{\mathrm{B}}+1)\sinh^2(r)+N_{\mathrm{B}}\cosh^2(r)\, \\
M&=&(2N_{\mathrm{B}}+1)e^{i\theta}\cosh(r)\sinh(r)\, .
\end{eqnarray*}
The master equation can be written as the sum of three Liouville operators
\begin{equation}\label{eq:Liouville}
\frac{d}{dt} \tilde \rho =(\mathcal{L}_{g}+\mathcal{L}_\Gamma+\mathcal{L}%
_\gamma)\tilde \rho\,.
\end{equation}
In the squeezed frame the first two terms correspond to the master equation
that is known from sideband cooling in ion traps (Eq.~\eqref{eq:IdealMaster}). Excitations of the qubit
on the red sideband followed by a spontaneous decay successively reduce the
phonon number and drive the oscillator into the ground state which
corresponds to a squeezed vacuum state in the original frame, $%
|\epsilon\rangle\!=\!U |0\rangle$. The third contribution, $\mathcal{L}%
_\gamma$ describes the coupling of the oscillator mode to a squeezed
reservoir~\cite{QuantumNoise}. The degree of squeezing of the reservoir is maximal since the
parameters $\tilde N$ and M are related by $|M|^2=\tilde N(\tilde N +1)$.
The different processes of the system dynamics in the squeezed frame are
summarized in Fig.~\ref{pic:LevelScheme2}. Note that $\tilde N$ and M grow
exponentially with the squeezing parameter while the Rabi frequency, $\tilde
g$ decreases exponentially. Therefore, even for a weak coupling, $\gamma$
the effects of the environment become essential as soon as r $\sim$ 1.

\emph{Weak coupling.} We first consider the limit where the decay of the
charge qubit is much faster than the rest of the system dynamics, $\Gamma
\gg \tilde g(r),\gamma \tilde N(r)$. In this regime the excited state can be
adiabatically eliminated by treating the coupling term, $\mathcal{L}_g$ in
second order perturbation theory. After tracing over the qubit degrees we
obtain a new master equation for the oscillator density operator, $\tilde \rho_x$.
The excitations of the charge qubit on the red sideband provide an
additional cooling rate of $4 \tilde g^2/\Gamma$ for the resonator. The master
equation for the oscillator alone can be solved by transforming it into a
partial differential equation for the Wigner function. The details of this
calculations are given in Appendix~\ref{app:Elimination}. In this limit
we obtain the result
\begin{equation}
\mathcal{R}=\sqrt{\frac{\frac{2\tilde g^2}{\Gamma}\,e^{-2r}+\frac{\gamma}{2}%
(2N_{B}+1)}{\frac{2\tilde g^2}{\Gamma}+\frac{\gamma}{2}}}\,.
\end{equation}
For small r and $\gamma$ this is close to the ideal behavior $\mathcal{R}=e^{-r}$
while for high values of r it saturates at the value of $\mathcal{R}=\sqrt{%
(2N_{B}+1)}$ which corresponds to the unperturbed thermal state.

\emph{Strong coupling.} In the strongly driven regime, $\tilde g\gg \Gamma,
\gamma \tilde N$, the system performs rapid oscillations between the states $%
|n,g\rangle$ and $|n-1,e\rangle$ on a timescale which is much faster than
the incoherent processes. It is therefore convenient to work in the basis of
dressed states $|\tilde n,\pm\rangle=\frac{1}{\sqrt{2}}(|n,0\rangle\pm
|n-1,1\rangle)$, $|\tilde 0\rangle=|0,0\rangle$. Since these are the
eigenstates of the Hamiltonian $H=\tilde g(a\sigma_+ +a^\dag \sigma_-)$ the
steady state density operator of Eq.~\eqref{eq:TransMaster} becomes diagonal
in that basis as $\tilde g\rightarrow \infty$. By neglecting the
off-diagonal terms the density operator can be approximated as
\begin{equation}
\tilde \rho= p_0|\tilde 0\rangle\langle \tilde 0|+\sum \, p_n^+|\tilde
n,+\rangle\langle \tilde n,+| +p_n^-|\tilde n,-\rangle\langle \tilde n,-|.
\end{equation}
Because the Liouville operators $\mathcal{L}_\Gamma$ and $\mathcal{L}_\gamma$
do not discriminate between the states $|\tilde n,\pm\rangle$ we define the
joint probabilities $p_n=p_n^+ +p_n^-$. With this ansatz the master equation~%
\eqref{eq:TransMaster} reduces to the rate equation
\begin{equation}  \label{eq:RateEquation}
\begin{split}
\dot p_n=&+ T_+(n-1)p_{n-1}+T_-(n+1) p_{n+1} \\
&-\left[ T_+(n)+T_-(n)\right]p_n\, ,
\end{split}%
\end{equation}
with the heating and cooling rates
\begin{equation}\label{eq:Rates}
\begin{split}
T_-(n)&=(\Gamma +\gamma (\tilde N+1)(2n-1))/2, \qquad T_-(0)=0\,, \\
T_+(n)&=\gamma \tilde N (2n+1)/2, \qquad T_+(0)=\gamma \tilde N\,. \\
\end{split}%
\end{equation}
In the stationary state the occupation numbers are determined by the
detailed balance condition $T_-(n+1)p_{n+1}=T_+(n)p_n$ and an analytic
expression for the mean occupation number is given by
\begin{equation}
\langle a^\dag a\rangle_{\tilde \rho}=\frac{\gamma \tilde N\left(2 \,_2F_1[\frac{3}{2},2,\frac{3%
}{2}+\alpha,z]-\, _2F_1[1,\frac{3}{2},\frac{3}{2}+\alpha,z]\right)}{%
\Gamma+\gamma(\tilde N +1)+ \gamma \tilde N\, _2F_1[1,\frac{3}{2},\frac{3}{2}%
+\alpha,z]} \, .
\end{equation}
$_2F_1$ denotes the hypergeometric function depending on the parameter $%
\alpha=\Gamma/(2\gamma(\tilde N +1))$ and is evaluated at the argument $%
z=\tilde N /(\tilde N +1)$. Since $\langle a^2\rangle_{\tilde \rho}$, $\langle a^{\dag
2}\rangle_{\tilde \rho} \rightarrow 0$ in the limit of strong coupling, a transformation
back to the original frame simply gives
\begin{equation}
\mathcal{R}=e^{-r}\sqrt{2\langle
a^\dag a \rangle_{\tilde \rho} +1}.
\end{equation}
\emph{Perturbation theory.} The interesting regime where the final state of
the resonator is squeezed, $\mathcal{R}<1$, obviously requires $\gamma \tilde{N}\ll \tilde g,\Gamma $%
. With this restriction a solution for arbitrary parameters $\tilde g$ and $\Gamma $
can be found by taking the ideal solution, $\tilde \rho =|0,g\rangle \langle 0,g|$
and treat the corrections of $\mathcal{L}_{\gamma }$ in first order
perturbation theory. The details of the calculations are listed in Appendix~%
\ref{app:Perturbation} and the result of this approach is
\begin{equation}
\mathcal{R}=\,e^{-r}\sqrt{1+\frac{\gamma \Gamma |M|}{2 \tilde g^{2}}+\gamma \tilde{N}%
\left( \frac{2}{\Gamma }+\frac{\Gamma }{2 \tilde g^{2}}\right) }\,.
\label{eq:PerturbationResult}
\end{equation}
While this expression gives the correct interpolation between the weak and the strong coupling limit
it is only valid for $\langle a^\dag a\rangle_{\tilde \rho}\ll 1$. A rough estimation shows that this is
still true up to the minimum, $\mathcal{R}_{\rm min}\!=\!{\rm min}\{\mathcal{R}(r),r>0\}$ in the case of $g_1 < \Gamma$
while expression~\eqref{eq:PerturbationResult} gives rather poor results for $\mathcal{R}_{\rm min}$ in the case of $g_1> \Gamma$.

\begin{figure}[tbp]
\begin{center}
\includegraphics[width=0.40\textwidth]{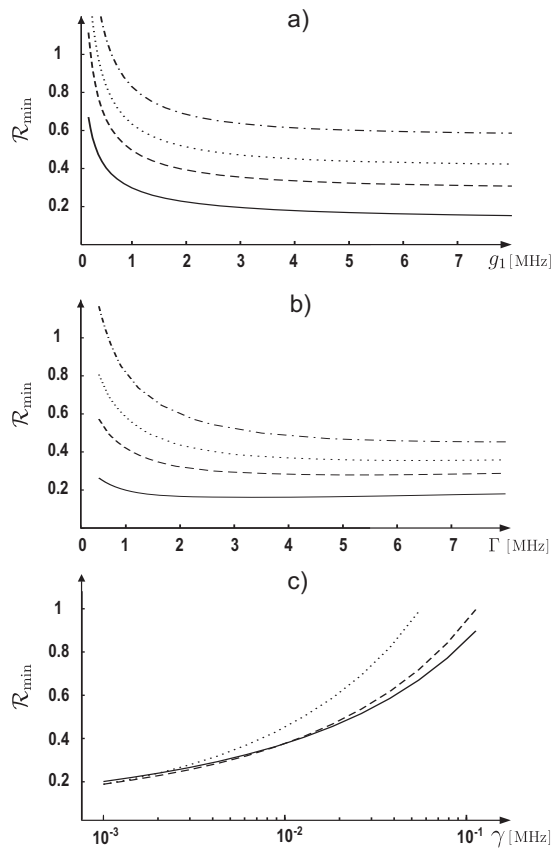}
\end{center}
\caption{Minimum values for the noise reduction, $%
\mathcal{R}_{\rm min}$ for various parameter values, $g_1$, $\Gamma$, $\gamma$ (all in MHz) and $N_B=5$. Figure a) shows the dependence
of $\mathcal{R}_{\rm min}$ on the driving strength, $g_1$ for a fixed $\Gamma=2$, while b) shows the dependence on the decay rate of the charge qubit,
$\Gamma$ for fixed $g_2=5$. The different values for $\protect\gamma$ in a) and b) are 0.001 (\emph{solid}), 0.005 (\emph{dashed}), 0.01 (\emph{dotted})
0.02 (\emph{dashed-dotted}). In Figure c) $\mathcal{R}_{\rm min}$ is plotted as a function of the phonon damping rate, $\gamma$ 
for $g_1=5$ and $\Gamma=2$
(\emph{solid}), $\Gamma=5$ (\emph{dashed}), $\Gamma=8$ (\emph{dotted}).}
\label{pic:Numerics}
\end{figure}

\emph{Numerical results.} For numerical calculations the master equation~%
\eqref{eq:TransMaster} is evaluated in the number basis. In the squeezed
frame the solution is close to the ground state so a relatively small number
of matrix elements is sufficient to describe the exact state. Fig.~\ref{pic:Numerics} shows
numerically calculated values of $\mathcal{R}_{\rm min}$ for various parameter values
for $g_1$, $\Gamma$ and $\gamma$. The results show that the noise, $\Delta X_1$ can be reduced to half of the standard quantum limit
for damping rates $\gamma \geq 0.02$. This corresponds to a Q-factor of 5000 in the case of a 100 MHz resonator. For $Q=10^5$ a
reduction by a factor of 5 is possible, still assuming a ``hot'' environment of about $T\approx 20 -30$ mK. Obviously, higher oscillator
frequencies or lower temperatures would improve the results even further.

%For a nanomechanical resonator with a fundamental frequency of $\nu=100$ MHz at an initial temperature of 30 mK. For $g_1=5$ MHz and
%$\Gamma=2$ MHz, a noise reduction by more than a factor of two can be reached with a quality factor of 10$^4$, while $\mathcal{R}=0.2$ is possible Q=10$^5$.

%%%%%%%%%%%%%%%%%%%%%%%%%%%%%%%%%%%%%%%%%%%%%%%%%%%%%%%%%%%%%%%%%%%%%%%%%%%%%5

%                             Detection

%%%%%%%%%%%%%%%%%%%%%%%%%%%%%%%%%%%%%%%%%%%%%%%%%%%%%%%%%%%%%%%%%%%%%%%%%%%%%%

\section{Detection of a squeezed state}

Below we discuss schemes for detecting non-classical states, in particular
squeezed states, of a nanomechanical system. The displacement measurements
based on laser interferometry, which are used for mechanical resonators with a length
in the order of a hundred microns, cannot be applied in the nanometer regime.
Alternatively displacement detectors based on a single electron transistor (SET)~\cite{Schoelkopf_Nature} have been considered~\cite{Blencowe_D,Hastings} 
and were recently used by the groups of A. Cleland~\cite{Cleland} and K. Schwab~\cite{Schwab_Science} 
to measure the fluctuation spectrum of nanomechanical resonators. 
While in current experiments the displacement sensitivity is still limited by the amplifier noise,
the quantum limit is determined by the back action of the current shot noise on the resonator.
By increasing the signal amplification of the detector, which is necessary to observe the reduced fluctuations of a squeezed state
also this back action is enhanced. A quantum mechanical analysis of the properties of SET-based displacement detectors 
is presented in~\cite{Hastings,Girvin_ShotNoise}. Using the results of the analysis done by Mozyrski \emph{et al.}~\cite{Hastings} 
the charge fluctuations on the SET island would destroy the squeezed state, 
especially in the experimentally attractive sequential tunneling regime.

In quantum optics with trapped ions and in Cavity QED experiments,
information about the oscillator state is often obtained via a coupling to a
two level atom. Efficient readout techniques for the state of an atom can be
used to measure properties of the not easily accessible oscillator mode.
Since CPB and other TLS are currently developed for quantum computation ~\cite{ShnirmanRev, Vion,Nakamura},
which in particular implies a read out of the qubit represented by e.g.~the
charge states of the CPB, this \textquotedblleft measurement
toolbox\textquotedblright\ is being developed in mesoscopic physics.
Motivated by this, we discuss the detection of the resonator state via the
readout of the charge qubit. We concentrate on two detection methods that can be performed with the setup shown in 
Fig.~\ref{pic:Setup} extended by a measuring device for the state of the CPB. 

\subsection{Dark Resonance} 
\begin{figure}[tbp]
\begin{center}
\includegraphics[width=0.45\textwidth]{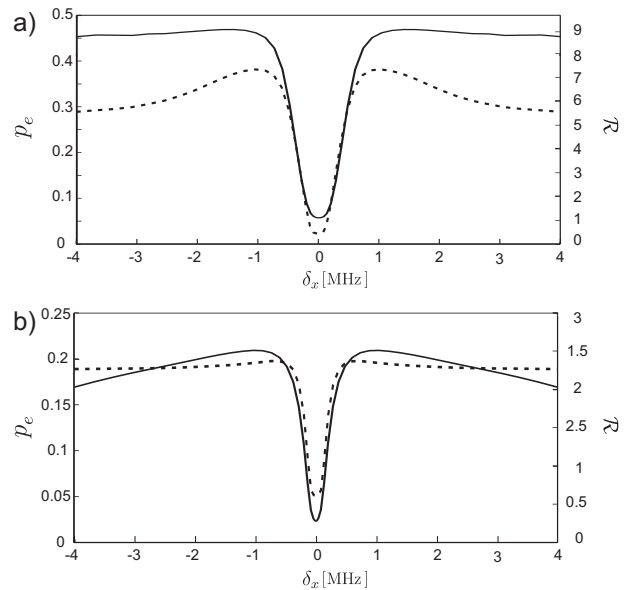}
\end{center}
\caption{Correspondence between the excited state population, $p_e$ (\emph{solid line}) and the degree of squeezing, $\mathcal{R}$
(\emph{dashed line}) as a function of the detuning, $\delta_x$. The values are plotted for the parameter values $\gamma=0.01$ MHz, $r=0.7$ and
$N_B=5$ for a) $g_1=5$ MHz, $\Gamma=2$ MHz (\emph{strong coupling}) and b) $g_1=1.5$ MHz, $\Gamma=5$ MHz (\emph{weak coupling}).}
\label{pic:Detuning}
\end{figure}

A simple way to verify the generation of a squeezed oscillator state is to look at the excited state
population, $p_e$ of the charge qubit. Since the squeezed state is a dark state of the system Hamiltonian~\eqref{eq:SqueezeHamiltonian}
the qubit excitations are significantly suppressed even in the presence of the driving fields. By varying the detuning $\delta_x$ a dark
resonance becomes visible at $\delta_x=0$~\cite{Zoller_Squeeze} which corresponds to the generation of a squeezed state. Fig.~\ref{pic:Detuning} shows
the expected correlations between the degree of squeezing and the steady state excitations of the qubit as a function of the detuning
$\delta_x$.

In the presence of a finite $\gamma$ the
population $p_e$ retains at a value of about $\gamma \tilde N/\Gamma\ll1$ (see Appendix~\ref{app:Perturbation}). In the regime of strong
coupling, $ g_1,g_2 \gg \Gamma$ this is clearly distinguishable from the value $p_e\approx 1/2$ as expected for, e.g. a thermal state.
For weak driving fields and a thermal oscillator state we expect a excited state population of
$p_e\approx \langle \hat n\rangle g_1^2/\Gamma^2+(\langle \hat n\rangle+1)g_2^2/\Gamma^2 $. Therefore the condition $g_2^2/\Gamma> \gamma
\tilde N$ to distinguish the squeezed state from a low temperature thermal state.

\subsection{Occupation Numbers}
According to its definition (Eq.~\eqref{eq:Definition}) a characteristic property of the (ideal) squeezed state,
$|\epsilon\rangle$ is that only the even 
number states are populated. The measurement of the resonator populations 
via the Stark shift of the qubit resonance frequency was proposed in Ref.~\cite{Schwab_Measure} while in Ref.~\cite{Santamore}
it is suggested to utilize the anharmonic coupling between bending modes for a Fock state readout.
Here we follow a different line~\cite{Ion_States} and use the linear coupling to the TLS to determine the
occupation numbers, $Q_n=\langle n|\rho_x|n\rangle$. 
The basic idea is as follows. Suppose we start from an initial density operator 
$\rho (0)=\rho _{x}\otimes |g\rangle \langle g|$ and switch on a Jaynes-Cummings coupling, $H_{\mathrm{I}%
}=\hbar g (a\sigma _{+}+a^{\dag }\sigma _{-})$ between the charge qubit and the resonator. 
The evolution of the qubit polarization is then given by
\begin{equation}
\langle \sigma _{z}\rangle (t)=-\sum_{n}Q_{n}\cos (2\Omega _{n}t)\,.
\label{eq:IdealSignal}
\end{equation}%
Due to the different Rabi frequencies $\Omega _{n}=g \sqrt{n}$ the
values of $Q_{n}$ can be extracted from the Fourier transform of
this signal.

In our system the required coupling of the oscillator to the qubit is realized 
by Hamiltonian~\eqref{eq:SqueezeHamiltonian} with $g_{1}=g$ and $%
g_{2}=0$. In contrast to the ideal situation of Eq.~\eqref{eq:IdealSignal}
the decay of the charge qubit and thermalization of the resonator mode lead to
modifications of the signal and restrict the applicability of this method.

Obviously, a necessary condition to resolve the oscillations of the qubit polarization
is the strong coupling regime $g\gg \Gamma ,\gamma (N_{B}+1)$%
. An approximate time evolution of the system can be obtained by the
following considerations. Starting from a pure state $|n,g\rangle $ the
system will oscillate between this state and $|n-1,e\rangle $ with a
frequency $2\Omega _{n}$, where $\Omega _{n}=g\sqrt{n}$. During this
oscillation it decays into neighboring number states with a rate $%
R_{n}\equiv T_{+}(n)+T_{-}(n)$ as defined in Eq.~\eqref{eq:Rates}.
Since all other states, $|m\not=n,g\rangle $ and $|m\not=n-1,e\rangle$ are
populated gradually their oscillations wash out and
with the exception of the ground state they give no contribution for $%
\sigma _{z}(t)$ . Therefore, for the
initial state $|n,g\rangle $ we obtain
\begin{equation}
\langle \sigma _{z}\rangle (t)=-p_{0}(n,t)-\cos (2\Omega
_{n}t)\,e^{-R_{n}t}\,,  \label{eq:ExpectedSignal}
\end{equation}%
where $p_{0}(t)$ is the population which accumulates in the ground state which has the form
\begin{equation}\label{eq:P0}
p_{0}(n,t)=1-e^{-\Gamma t/2}\sum_{k=0}^{n-1}\frac{1}{k!}\left(\frac{\Gamma t}{2}\right)^k\,,
\end{equation}
in the limit $\Gamma \gg \gamma(N_B+1)$. The exact time dependence for $\gamma>0$ is not important since
the changes of $p_0(n,t)$ are slow compared to the Rabi oscillations.
For an arbitrary initial state with occupation numbers $Q_{n}$ the
polarization of the qubit is given by
\begin{equation}
\langle \sigma _{z}\rangle (t)=-\sum_{n}Q_{n}\left( p_{0}(n,t)+\cos (2\Omega
_{n}t)\,e^{-R_{n}t}\right) \,.  \label{eq:analyticsz}
\end{equation}%
The extraction of the occupation probabilities from a given function $%
\langle \sigma _{z}\rangle (t)$ requires the resolution of the individual
Lorentzian peaks in the Fourier transform of this signal. The $n-$th peak
can be resolved if the condition $R_{n}+R_{n+1}<g/\sqrt{n}$ is true.
Therefore, a lower bound for the maximum occupation probability which we can
determine with this method is given by the solution of the equation
\begin{equation}
(\Gamma +2\gamma N_{B})n_{\mathrm{max}}^{1/2}+2\gamma (2N_{B}+1)n_{\mathrm{%
max}}^{3/2}=g\,.
\end{equation}%
A first order approximation, valid for $\gamma g^{2}/\Gamma ^{2}\ll 1$ gives
\begin{equation}
n_{\mathrm{max}}\simeq \frac{g^{2}}{\Gamma ^{2}}\left( 1-\frac{4\gamma
(2N_{B}+1)g^{2}}{\Gamma ^{3}+6\gamma (2N_{B}+1)g^{2}}\right) \,.
\end{equation}%
For the parameter values $g=5\,\Gamma $, $N_{B}=5$ and $\gamma
=0.001\,-\,0.01\,\Gamma $ we obtain $n_{\mathrm{max}}\approx 8-17$, which is
sufficient to detect slightly squeezed or low number states.

\begin{figure}[tbp]
\begin{center}
\includegraphics[width=0.45\textwidth]{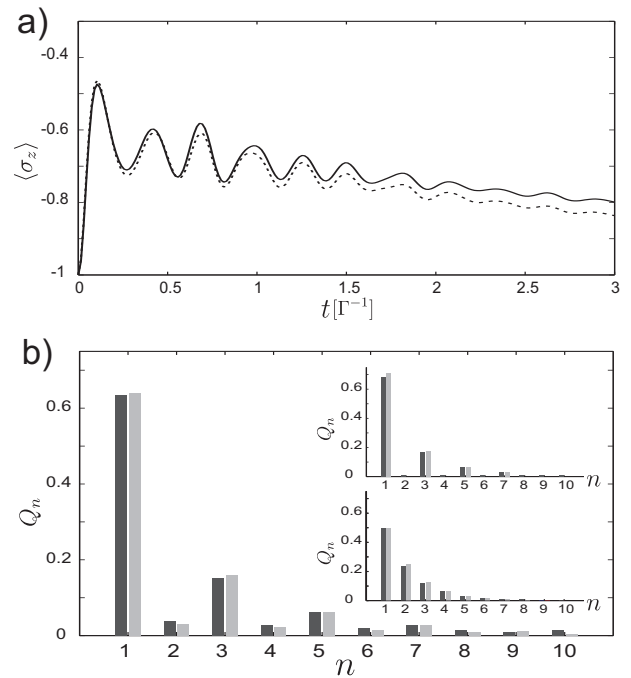}
\end{center}
\caption{Measurement of the oscillator occupation numbers, $Q_n$. A squeezed state is generated as explained in
the previous sections with (all in MHz) $g_1=5$, $g_2=3.5$ ($r\!=\!0.88$), $\Gamma=2$, $\gamma=0.01$ and $N_B=5$ MHz. For the detection 
we assume the same values but $g\!=\!g_1\!=\!8$ and $g_2=0$. Figure a) shows the exact time evolution of $\langle \sigma_z\rangle(t)$ (\emph{solid})
which is compared to the expected behavior (\emph{dashed}) as given in Eq.~\eqref{eq:ExpectedSignal}. The difference for larger times is due to the
deviations of the exact time dependence of $p_0(n,t)$ from Eq.~\eqref{eq:P0} for $\gamma>0$. In Figure b) the extracted occupation probabilities
(\emph{dark gray}) 
are compared with the exact values (\emph{light gray}). The inset shows the estimated number distribution for a ideal squeezed state 
(\emph{top}) and a thermal state with the 
same mean occupation number, $\langle n \rangle =1$ (\emph{bottom}). }
\label{pic:Populations}
\end{figure}

Fig.~\ref{pic:Populations} shows the time evolution of $\langle \sigma_z\rangle(t)$ and the extracted occupation
probabilities, $Q_n$ for a non-ideal squeezed state. A clear distinction between the squeezed state and a
thermal state with the same mean occupation number is possible.

Although this method provides a complete determination of the number state populations for states close to the
ground state, it is not possible to distinguish between coherent superpositions and mixed states. Therefore, in 
the next section, we discuss the implementation of quantum state tomography to 
obtain the full information about the resonator's density matrix.

%%%%%%%%%%%%%%%%%%%%%%%%%%%%%%%%%%%

%%%%%%%%         State Tomography

%%%%%%%%%%%%%%%%%%%%%%%%%%%%%%%%%%%

\section{Quantum State Tomography}\label{sec:Tomography}

The ultimate determination of an arbitrary quantum state is the measurement of the
complete density operator. The procedure of estimating the density matrix by repeated measurements
on the same initial state is called \emph{quantum tomography}~\cite{DAriano}. For a harmonic oscillator the Wigner function of a state,
a quasi probability distribution in phase space~\cite{QuantumNoise,Schleich}, contains the same information as the density matrix.
The implementation of a method to reconstruct the Wigner function for a nanomechanical resonator
provides an universal tool to detect and characterize
various non-classical states, and therefore a tool to clearly demonstrate the
quantum nature of this still macroscopic system.

In quantum optics methods for state tomography are well known~\cite{Vogel,Davidovich,Schleich} and have been
successfully implemented to detect non-classical states of a cavity field~\cite{CavityTomography} or
the motion of a trapped ion~\cite{Leibfried}. We consider the method discussed in Ref.~\cite{Leibfried} and 
show that it is appropriate for the implementation in nano scale mechanical systems.

To determine the Wigner function at a certain point in phase space, $\alpha
=x+ip$ we start from the identity~\cite{Schleich}
\begin{equation}
W(\alpha )=\frac{2}{\pi}\sum_{n=0}^{\infty }(-1)^{n}Q_{n}(\alpha )\,,  \label{eq:Wigner1}
\end{equation}%
where $Q_{n}(\alpha )=\langle n|D^{\dag }(\alpha )\rho _{x}D(\alpha
)|n\rangle $ are the occupation probabilities of the displaced density
operator. The values of $Q_{n}(\alpha )$ are measured as discussed in the previous section.

A complete state tomography consists of the following steps. For each point
in phase space $(x,p)$ we displace the original density operator by $\alpha
=-(x+ip)$. Then the occupation numbers of the oscillator are determined by
measuring the time evolution of the qubit polarization, $\langle \sigma _{z}\rangle (t)$.
In the end we obtain the measured value of the Wigner function, $\tilde W(\alpha)$ by 
summing up the $Q_n(\alpha)$ according to
Eq.~\eqref{eq:Wigner1}.

In the following we discuss the limits for the implementation of the individual steps 
for a system consisting of a
nanomechanical resonator coupled to a charge qubit.

\emph{Applying the displacement operator.}
The first step of the procedure described above
requires the shift of the oscillator by the complex amplitude $\alpha $. In
the setup shown in Fig.~\ref{pic:Setup} this displacement can be achieved
either by applying an additional voltage to a lead opposite the CPB or by
exploiting the existing coupling to the charge qubit, $\lambda (t)(a+a^{\dag })\sigma _{x}$.

In the first case a voltage drop $V_{d}$ over a capacitance $C_{d}(x)$
formed by the lead and the resonator generates the driving Hamiltoninan, $%
H_{d}=\lambda (t)(a+a^{\dag })$ with $\lambda =\frac{1}{2}%
C_{d}V_{d}^{2}x_{0}/d$. The evolution of the oscillator under $H=\hbar \nu
a^{\dag }a+H_{d}$ is~\cite{QuantumNoise}
\begin{equation}
|\psi (t)\rangle =\hat{D}(\alpha (t))e^{-i\nu ta^{\dag }a}|\psi (0)\rangle
\,,
\end{equation}%
with
\begin{equation}
\alpha (t)=-i\int_{0}^{t}\,dt^{\prime }\,e^{i\nu (t-t^{\prime })}\lambda
(t^{\prime })
\end{equation}%
Because $\lambda \sim 1$ GHz $\gg \nu $, a short constant voltage pulse is
sufficient to obtain displacements of $|\alpha |\leq 10$.

If the coupling to the CPB is used for the displacement, we first transfer
the ground state of the qubit $|0\rangle $ into one of the eigenstates of $%
\sigma _{x}$ by adiabatically changing the CPB parameters $V_{g}^{0}$ and $E_J$. Since
the coupling strength, $\lambda \approx $ 5-10 MHz is much lower than the
oscillator frequency a radio frequency pulse $\lambda (t)\sim \cos (\nu t)$
has to be applied to achieve shifts in the order of $|\alpha |\geq 1$.

Especially in the second case the Wigner function of the oscillator is
modified during the displacement due to thermalization with the phonon bath.
The diffusive dynamics of a driven oscillator and therefore the resulting errors
can be calculated exactly (see Appendix~\ref{app:Displacement}).
If we assume an initial Gaussian distribution and a displacement time $\Delta
t=|\alpha |/|\lambda |\ll 1/\gamma $ we obtain the relative errors for the
widths
\begin{equation}
\varepsilon (\Delta _{x,p}^{2})=\frac{\gamma |\alpha |}{|\lambda |}\left( 1+%
\frac{N_{B}+\frac{1}{2}}{2\Delta _{x,p}^{2}}\right) \,.
\label{eq:ErrorDelta}
\end{equation}%
The damping also modifies the displacement amplitude. Together with
the deviation caused by a small detuning in the driving frequency $\omega
=\nu +\delta _{x}$, we obtain a relative error
\begin{equation}
\varepsilon (|\alpha| )=\frac{|\alpha |}{|\lambda |}\left( \frac{|\delta _{x}|%
}{2}+\frac{\gamma }{4}\right) \,.
\end{equation}%
For the parameter values considered in this paper $\varepsilon (|\alpha| )\sim
10^{-3}$.

\emph{Occupation numbers.}
The measurement of the probabilities, $Q_n(\alpha)$ is done as discussed in the previous section. 

\begin{figure}[tbp]
\begin{center}
\includegraphics[width=0.4\textwidth]{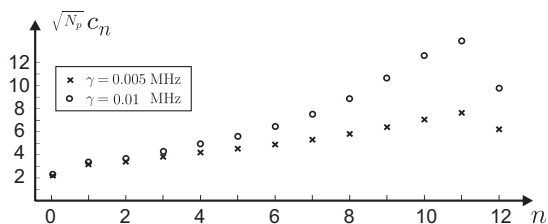}
\end{center}
\caption{Coefficients, $c_n$ which determine the error of the estimated
occupation numbers according to Eq.~\eqref{eq:RealError}. For a large number of measured data points, $N_D$
the $c_n$ are proportional to $1/\sqrt{N_D}$. See text for details.}
\label{pic:Cn}
\end{figure}

For an estimation of the error due to truncation we consider an initially density
matrix of a pure number state, $\rho =|m\rangle \langle m|$ with $m<n_{%
\mathrm{max}}$. The application of the displacement operator shifts some
part of the wavefunction out of the detectable subspace $\{|0\rangle ,\dots
,|n_{\mathrm{max}}\rangle \}$. This leads to an absolute error in the
estimated Wigner function $\tilde{W}(\alpha )$ of
\begin{equation}
\varepsilon (m,\alpha )=|\tilde{W}(\alpha )-W(\alpha )|\leq\frac{2}{\pi} \sum_{k=n_{%
\mathrm{max}}}^{\infty }|\langle k|\hat{D}(-\alpha )|m\rangle |^{2}\,.
\label{eq:ErrorEst}
\end{equation}%
The matrix elements of the displacement operator are given by
\begin{equation*}
|\langle k|\hat{D}(-\alpha )|m\rangle |^{2}=\frac{m!}{k!}|\alpha
|^{2(m-k)}e^{-|\alpha |^{2}}\left[ L_{m}^{k-m}(|\alpha |^{2})\right] ^{2}\,
\end{equation*}%
where $L_{m}^{k-m}$ are the generalized Laguerre polynomials. The sum $%
\sum_{m}\langle m|\rho |m\rangle \,\varepsilon (m,\alpha )\ll 1$ provides an upper bound
for the total truncation error and rough estimation whether the method is applicable or not.
In practice the determination of the $Q_{n}$ is done by fitting the actual signal by minimizing
the deviations in the least square sense. For a linear fit the standard deviation of the $Q_{n}$
can be written as
\begin{equation}
\sigma (Q_{n})=c_{n}\sigma _{D}\,.  \label{eq:RealError}
\end{equation}%
The error in the measured data point, $\sigma _{D}$ includes the error from
the measurement of the qubit polarization as well as deviations of the
system evolution from expected evolution as given in Eq.~\eqref{eq:ExpectedSignal}.
The coefficients, $c_{n}$ depend on the parameters of the system and the
fitting procedure~\cite{NumRecipes}. While the $c_n$ increase on a scale set by $n_{\rm max}$ it is still
possible to determine the $Q_n$ beyond $n_{\rm max}$ in the expense of accuracy.
For the example given below the $c_{n}$ are shown in Fig.~%
\ref{pic:Cn}.

\emph{Example.} To summarize the considerations made in this section we discuss the
generation of a squeezed state and the reconstruction of its Wigner function
for a specific example in some detail. We consider a nanomechanical
resonator with a fundamental frequency of $100$ MHz and a $Q$-value of $%
2\times10^{4}$ cooled to a temperature of $25$ mK. For the coupling to the CPB we
assume the values $g_{1}=5$ MHz, $g_{2}=3$ MHz ($r=0.7$) and a decay rate $%
\Gamma =2$ MHz. Using the results of the last section, this drives the
resonator into a state with a noise reduction $\mathcal{R}\approx 0.5$ in
one quadrature components. After the resonator has reached its steady state
we switch off the coupling and wait for a short time $\tau \sim 1/\Gamma $
to reduce the excited state population to less than 1\%. According to Eq.~
\eqref{eq:ErrorDelta} this delay leads to a broadening in the $X_1$ direction
of about 17\%, so we end up with $\mathcal{R}\approx $ $0.6$. The exact Wigner function 
for this state is well located in a the phase space region $|\alpha|\leq 2$.

For the displacement of the oscillator we use the coupling to the CPB as
described above with $\lambda =8$ MHz. Eq.~%
\eqref{eq:ErrorDelta} predicts an error of $\varepsilon(\Delta_x^2) \approx |\alpha
|\times $2\%

For the determination of the occupation numbers, $Q_{n}(\alpha )$ we apply a
red sideband signal with $g=8$ MHz. For this value we obtain $n_{\mathrm{max}%
}\approx 8$. The estimation $\sum_{m}\langle m|\rho |m\rangle \,\varepsilon
(m,\alpha )\leq 0.15$ for the phase space region $|\alpha |\leq 2$ shows
that an accurate reconstruction of the Wigner function is possible. To
resolve oscillations of frequency $g\sqrt{n_{\mathrm{max}}}$ over a time
much longer than the characteristic decay time the number of measured data
point, $N_{D}$ has to fulfill $N_{D}\gg 2g\sqrt{n_{\mathrm{max}}}/(\pi
\Gamma )$. In our example we choose a measurement time $T=3$ $\Gamma^{-1} $ and $%
N_{D}=150$. For these values the coefficients $c_{n}$ in Eq.~%
\eqref{eq:RealError} are plotted in Fig.~\ref{pic:Cn}. We suppose that the qubit
polarization can be measured with an accuracy of $0.02$ and that the values for $%
Q_{0\dots 12}$ (we choose $n_{\rm max}=12$) are estimated. Adding the errors of the occupation numbers,
the error due to truncation and the error from the displacement we expect an accuracy for the reconstructed
Wigner function of
$|\tilde{W}(\alpha )-W(\alpha )| \leq 0.05$.

The results of a numerical simulation of the generation of the squeezed state and the reconstruction of
its Wigner function is shown in Fig.~\ref{pic:StateTomography}. 

In Appendix~\ref{app:Tomography} we also briefly discuss a different method for a state tomography as proposed by Lutterbach 
and Davidovich~\cite{Davidovich}. While this method is experimentally more attractive since it requires less measurements, a stronger 
coupling and longer decoherence times of the qubit are required for its implementation.
 
\begin{figure}[tbp]
\begin{center}
\includegraphics[width=0.40\textwidth]{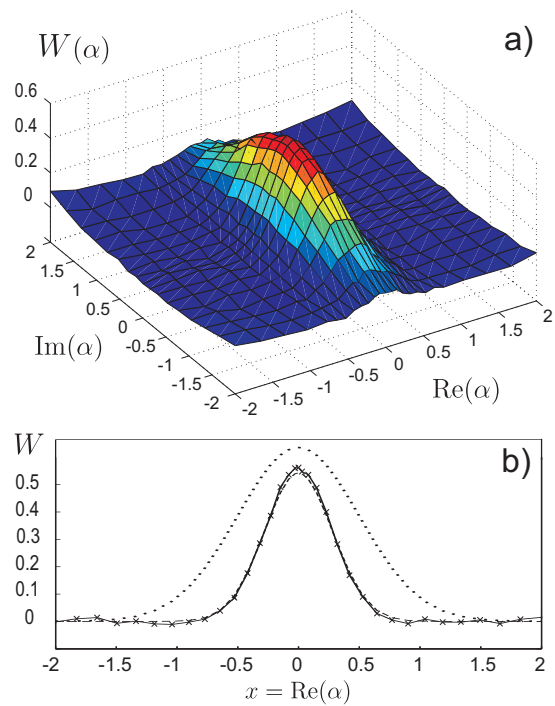}
\end{center}
\caption{Results of the numerical simulation of the state tomography for a squeezed state as described in Section~\ref{sec:Tomography}.
Figure a) shows the reconstructed Wigner function, $\tilde W(\alpha)$.
In Figure b) a cut of $\tilde W(\alpha)$ (\emph{solid line}) in x direction (${\rm Im}(\alpha)=0 $)
is compared with the corresponding values of the real Wigner function, $W(\alpha)$ (\emph{dashed line})
and the Wigner function of the oscillator ground state (\emph{dotted line}).}
\label{pic:StateTomography}
\end{figure}

\section{Conclusion}
In this paper we showed that by \emph{reservoir engineering} the fundamental mode of a nanomechanical resonator can be driven into 
a squeezed state. The stationary state exhibits noise reduction in one of the quadrature components by a factor of 
$\mathcal{R}\approx$ 0.5 - 0.2 below the standard
quantum limit. For a 100 MHz resonator these values are obtained for Q-values in the order of $Q=10^4-10^5$ and standard dilution refrigerator
temperatures of T $\approx$ 30 mK. The detection of the squeezed state can be done within the same setup as used for its generation by measuring 
the excitation probability of the charge qubit. Furthermore this measurement procedure can be extended to obtain a
complete reconstruction of the oscillator's Wigner function. This tool
provides an universal detection method for non-classical behavior of the resonator.

\begin{acknowledgments}
The authors thank L. Tian, I. Wilson-Rae and A. Imamoglu for useful discussions. P.R.
also thanks G. Sch\"on, Y. Makhlin and G. Johansson for hospitality and fruitful discussions.
Work at Innsbruck was supported in part by the Austrian Science Foundation FWF, European
Networks and the Institute for Quantum Information.
\end{acknowledgments}

%%%%%%%%%%%%%%%%%%%%%%%%%%%%%%%%%%%%%%%%%%%%%%%%%%%%%%%%%%%%%%%%%%%%%%%%%%

%%%%%%%%%%%%%%%                 Appendix

%%%%%%%%%%%%%%%%%%%%%%%%%%%%%%%%%%%%%%%%%%%%%%%%%%%%%%%%%%%%%%%%%%%%%%%%%

\begin{appendix}
\section{Adiabatic Elimination}\label{app:Elimination}
We consider master equation \eqref{eq:Liouville} in the limit
$\Gamma \gg \tilde g, \gamma \tilde N$. To zero order in $\tilde g/\Gamma, \gamma/\Gamma$
the equation reduces to $\di_t\tilde \rho=\mathcal{L}_\Gamma\tilde \rho$ and a
steady state solution is located within
the subspace $S=\{\tilde \rho_x\otimes |g\rangle\langle g|\}$. The term
$\mathcal{L}_{\gamma}$ leads to a slow dynamic within this
subspace while $\mathcal{L}_g$ couples S to its complement. Since $\mathcal{L}_\Gamma \tilde \rho =\mathcal{O}(\Gamma)$, $\forall \rho \notin S$ we can
project the master equation~\eqref{eq:TransMaster} on the subspace
S and treat the coupling of $\mathcal{L}_g$ in perturbation
theory. After tracing over the charge qubit states we obtain
\begin{equation}
\frac{d\tilde \rho_x}{dt}=\mathcal{L}_\gamma \tilde \rho_x
-\mathrm{Tr}_{CQ}\{\mathcal{L}_g
\mathcal{L}_\Gamma^{-1}\mathcal{L}_g \,\tilde \rho_x\!\otimes\!|g\rangle\langle
g|\}\,.
\end{equation}
An evaluation of this expression leads to a master equation
for the reduced density operator of the resonator, $\tilde \rho_x$ given by
\begin{equation}
\begin{split}
\frac{d\tilde \rho_x}{dt}=&\left(\frac{2 \tilde g^2}{\Gamma}+\frac{\gamma}{2}(\tilde N +1)\right)
                           \left(2a\tilde\rho_x a^\dag-a^\dag a\tilde\rho_x-\tilde\rho_x a^\dag a \right)\\
          &+\frac{\gamma}{2}\tilde N\left(2a^\dag\tilde\rho_x a- a a^\dag \tilde\rho_x - \tilde\rho_x a a^\dag\right)\\
      &-\frac{\gamma}{2}M\left(2a\tilde\rho_x a-a a\tilde\rho_x-\tilde\rho_x a a \right)\\
          &-\frac{\gamma}{2}M^*\left(2a^\dag\tilde\rho_x a^\dag-a^\dag a^\dag\tilde\rho_x-\tilde\rho_x a^\dag a^\dag \right) \, .
\end{split}
\end{equation}
Because the master equation now only contains the operators a and
$a^\dag$  it can be transformed into a partial differential equation for
the Wigner function \cite{Walls},
\begin{equation}\label{eq:DiffWigner}
\begin{split}
\frac{\di}{\di t}
\tilde W(\alpha)=&\left(\frac{2 \tilde g^2}{\Gamma}+\frac{\gamma}{2}\right)\left(\frac{\di}{\di
\alpha}\alpha
+\frac{\di}{\di \alpha^*}\alpha^* \right)\tilde W(\alpha)\\
                        +&\left(\frac{2g^2}{\Gamma}+\frac{\gamma}{2} (2\tilde N+1) \right)\frac{\di^2}{\di \alpha \di\alpha^*}\tilde W(\alpha)\\
            +&\frac{\gamma}{2}\left(M\frac{\di^2}{\di \alpha^2}+M^*\frac{\di^2}{\di \alpha^{*2}}\right)\tilde W(\alpha) \, .
\end{split}
\end{equation}
By introducing the new variables $\tilde x$, $\tilde p$ defined by
$\alpha= (\tilde x +i \tilde p)e^{i\theta/2} $,
Eq.~\eqref{eq:DiffWigner} separates into two independent
Fokker-Planck equations and their steady state solution is
\begin{equation}
\tilde W(\tilde x,\tilde p)=\mathcal{N}\exp\left(-\left( \frac{2 \tilde g^2}{\Gamma}+\frac{\gamma}{2}\right)
\left(\frac{\tilde
x^2}{D_x}+\frac{\tilde p^2}{D_p}\right)\right)\,,
\end{equation}
where $\mathcal{N}$ is a normalization constant and
\begin{eqnarray*}
 D_{x,p}=\frac{\tilde g^2}{\Gamma}+\frac{\gamma}{2}\left(N_B+\frac{1}{2}\right)e^{\pm 2 r} \,.
\end{eqnarray*}
Because these calculations have been performed in the squeezed
frame we need another change of variables to undo this
transformation and finally obtain the Wigner function in the
original frame,
\begin{equation}
W(x,p)=\mathcal{N}\exp\left(-\frac{1}{2}(x,p)\Sigma^{-1}(x,p)^T\right)\,,
\end{equation}
where the variance matrix $\Sigma$ is
\begin{equation}
\Sigma^{-1}= \mathbf{T}^T
            \left( \begin{array}{cc} k/D_x & 0 \\ 0 & k/D_p \end{array}\right) \mathbf{T}\, ,
\end{equation}
with
\begin{equation*}
\mathbf{T}\!=\!\left( \begin{array}{cc}
e^r & 0 \\
0 & e^{-r}
\end{array}\right)
\left( \begin{array}{cc} \cos(\theta/2) &
\sin(\theta/2) \\ -\sin(\theta/2) & \cos(\theta/2)
\end{array}\right) \, .
\end{equation*}
The variances of the two quadrature components, $X_{1,2}$ are given by the widths of the Gaussian
function $W(X_1,X_2)$,
\begin{equation}
\begin{split}
(\Delta X_1)^2=&\frac{1}{4}\left(\frac{2g^2}{\Gamma}\,e^{-2r}+\frac{\gamma}{2}(2N_B+1)\right)/\left(\frac{2g^2}{\Gamma}+\frac{\gamma}{2}\right)\,,\\
(\Delta X_2)^2=&\frac{1}{4}\left(\frac{2g^2}{\Gamma}\,e^{2r}+\frac{\gamma}{2}(2N_B+1)\right)/\left(\frac{2g^2}{\Gamma}+\frac{\gamma}{2}\right)\,.
\end{split}
\end{equation}

\section{Perturbative Solution}\label{app:Perturbation}
For a weak coupling of the oscillator to the phonon bath we write
the master equation \eqref{eq:TransMaster} as
\begin{equation}
\frac{d\tilde \rho}{d t}=\mathcal{L}_{0}(\tilde \rho)+\mathcal{L}_{\gamma}(\tilde \rho)
\end{equation}
where $\mathcal{L}_0=\mathcal{L}_g+\mathcal{L}_\Gamma$ contains
the terms proportional to $\tilde g$ and $\Gamma$ with the steady state
solution $\tilde \rho_0=|0\rangle\langle 0|\otimes|g\rangle\langle g|$.
In the limit of weak dissipation, $\gamma (\tilde N+1) \ll \tilde g, \Gamma$
we write the density operator as $\tilde \rho=\tilde \rho_{0}+\tilde \rho_1$, and
calculate the corrections, $\tilde \rho_1$ up to first order in $\gamma$.
\begin{equation}
\begin{split}
\frac{d \tilde \rho_1}{d t}=
%\di_t \,\tilde  \rho_1 =
&\,\mathcal{L}_{0}(\tilde \rho_1)+\mathcal{L}_\gamma(\tilde \rho_0)\\
   =&-i[\,\tilde g(a\sigma_+ +a^\dag\sigma_-)-i\frac{\Gamma}{2}|e\rangle\langle e|,\tilde \rho_1\,]+\Gamma\sigma_{-}\tilde \rho_1\sigma_{+}\\
            &+\gamma \tilde N (|1\rangle\langle 1|\!-\!|0\rangle\langle 0|)+\gamma M|0\rangle\langle 2|+\gamma M^*|2\rangle\langle 0|\,.
\end{split}
\end{equation}
The steady state solution can be calculated e.g. by evaluating this equation in the number
state basis. Using the notation $P_n=\langle
g, n+1|\tilde \rho_1|e,n\rangle- \langle e,n|\tilde \rho_1|g,n+1\rangle$,
$E_n=\langle e,n|\tilde \rho_1|e,n\rangle$ and $G_n=\langle
g,n|\tilde \rho_1|g,n\rangle$, we obtain the set of coupled equations
\begin{eqnarray*}
-i\tilde g\sqrt{n+1}P_n-\Gamma E_n&=&0\, ,\\
-\Gamma/2 P_n-i2\tilde g\sqrt{n+1}(E_n-G_{n+1})&=&0 \, ,\\
i\tilde g\sqrt{n}P_{n-1}+\Gamma E_n+\gamma \tilde N \delta_{n,1}-\gamma
\tilde N \delta_{n,0}&=& 0\, .
\end{eqnarray*}
Because all matrix elements for $n>1$ are zero we find a very
simple solution for the mean occupation number and the excited state  population
\begin{equation}\label{eq:PerturbMeanNumber}
\begin{split}
\langle\hat n\rangle_{\tilde \rho}=G_1+E_1&=\gamma \tilde
N\left(\frac{1}{\Gamma}+\frac{\Gamma}{4\tilde g^2}\right) \,,\\
\langle |e\rangle\langle e| \rangle_{\tilde \rho}&=E_0=\frac{\gamma\tilde N}{\Gamma}\,.\\
\end{split}
\end{equation}
Another set of equations for the matrix elements between the
states $\langle n|$ and $|n+2\rangle$ gives two non-zero
contributions for $\langle g,0|\rho_1|g,2\rangle$ and $\langle
g,2|\rho_1|g,0\rangle$ which lead to
\begin{equation}
\langle a^2\rangle_{\tilde \rho}+\langle a^{\dag 2}
\rangle_{\tilde \rho}=\frac{\gamma\Gamma}{4\tilde g^2}(M+M^*)\, .
\end{equation}
The variance of the $X_1$ quadrature component in the original frame is just given by
\begin{equation}
(\Delta X_1)^2=\frac{e^{-2r}}{4}\left(\langle a^2\rangle_{\tilde \rho}+\langle a^{\dag 2}\rangle_{\tilde \rho}+
2\langle\hat n\rangle_{\tilde \rho}+1\right)\, .
\end{equation}
This leads to the result of Eq.~\eqref{eq:PerturbationResult}.

\section{Dissipative, driven Oscillator}

\label{app:Displacement} We consider a harmonic oscillator, $H=\hbar \nu
a^{\dag }a$ driven by the linear term $H_{d}=\lambda e^{i\nu t}a+\lambda
^{\ast }e^{-i\nu t}a^{\dag }$ which is weakly coupled to a reservoir. The
master equation in the interaction picture is given by
\begin{equation}
\begin{split}
\frac{d\rho }{dt}=-i[\lambda a+\lambda a^{\dag },\rho ]& +\frac{\gamma (N_{%
\mathrm{B}}\!+\!1)}{2}\left( 2a\rho a^{\dag }\!-\!a^{\dag }a\rho \!-\!\rho
a^{\dag }a\right) \\
& +\frac{\gamma N_{\mathrm{B}}}{2}\left( 2a^{\dag }\rho a\!-\!aa^{\dag }\rho
\!-\!\rho aa^{\dag }\right) \,.
\end{split}%
\end{equation}%
This equation can be transformed into a Fokker-Planck equation for the
Wigner function~\cite{Walls}. For the coordinates $x=(\alpha +\alpha ^{\ast
})/2$ and $p=(\alpha -\alpha ^{\ast })/2i$ we obtain
\begin{equation}
\begin{split}
\frac{dW}{dt}& =\left[ \mathrm{Re}(\lambda )\frac{\partial }{\partial p}+%
\mathrm{Im}(\lambda )\frac{\partial }{\partial x}\right] W \\
& +\frac{\gamma }{2}\left[ \left( \frac{\partial }{\partial x}x+\frac{%
\partial }{\partial p}p\right) +\frac{N_{B}+\frac{1}{2}}{2}\left( \frac{%
\partial ^{2}}{\partial x^{2}}+\frac{\partial ^{2}}{\partial p^{2}}\right) %
\right] W.
\end{split}%
\end{equation}%
A general solution for an initial distribution, $W_{i}(\alpha )$ is given by
\begin{equation}
W(\alpha ,t)\!=\!\frac{1}{{2\pi \sigma ^{2}(t)}}\int \! d^{2}\alpha_0
\,W_{i}(\alpha _{0})\,\mathrm{exp}\left( \!-\frac{|\alpha -\alpha _{0}e^{-%
\frac{\gamma t}{2}}|^{2}}{2\sigma ^{2}(t)}\right)
\end{equation}%
with $2\sigma ^{2}(t)=(N_{B}+\frac{1}{2})(1-e^{-\gamma t})$. For the special case where the
the initial distribution is a Gaussian function centered at the origin the time dependent solution
can be evaluated as
\begin{equation}
W(x,p,t)=\mathcal{N}(t)\,\mathrm{exp}\left( -\frac{(x-c_{x}(t))^{2}}{2\Delta
x^{2}(t)}-\frac{(p-c_{p}(t))^{2}}{2\Delta p^{2}(t)}\right) \, ,
\end{equation}%
with a normalization factor, $\mathcal{N}(t)$ and the time dependent parameters
\begin{equation}
\begin{split}
\Delta x^{2}(t)& =\Delta x^{2}(0)e^{-\gamma t}+\frac{2N_{B}+1}{4}%
(1-e^{-\gamma t})\,, \\
\Delta p^{2}(t)& =\Delta p^{2}(0)e^{-\gamma t}+\frac{2N_{B}+1}{4}%
(1-e^{-\gamma t})\,, \\
c_{x}(t)& =2\,\mathrm{Im}(\lambda )(1-e^{-\gamma t/2})/\gamma \,, \\
c_{p}(t)& =2\,\mathrm{Re}(\lambda )(1-e^{-\gamma t/2})/\gamma \,.
\end{split}%
\end{equation}

\section{Fast Tomography}\label{app:Tomography}
The reconstruction of the Wigner function as described in Section~\ref{sec:Tomography} requires a lot of measurements since it relies 
on the whole time evolution of the qubit polarization. A direct measurement of the Winger function form a single value of $\langle \sigma_z \rangle$
has been proposed by Lutterbach and Davidovich~\cite{Davidovich} for Cavity QED and ion traps.
Here we give a brief summary of this procedure which relies on the identity
\begin{equation}
W(\alpha )=\frac{2}{\pi}\mathrm{{Tr}\{\hat{D}^{\dag }(\alpha )\rho_x \hat{D}(\alpha
)e^{i\pi a^{\dag }a}\}\,.}  \label{eq:Wigner}
\end{equation}%
Again the measurements on the qubit are used to deduce the expectation
value of the parity operator, $e^{i\pi a^{\dag }a}$ . To do so we assume that a certain
evolution $U$ can be applied to the system with the following properties:
\begin{eqnarray}
U(|g\rangle |n\rangle ) &=&|g\rangle |n\rangle \qquad \mathrm{{%
if\,\,n\,\,is\,\,odd}\,,}  \notag  \label{eq:U} \\
U(|g\rangle |n\rangle ) &=&|e\rangle |n\rangle \qquad \mathrm{{%
if\,\,n\,\,is\,\,even}\,.}
\end{eqnarray}%
Starting from an initial density operator $\rho (0)=\rho_x \otimes |0\rangle
\langle 0|$ the Wigner function at the point $\alpha $ can then be
reconstructed in three steps: First, apply the displacement operator, $%
D(-\alpha )=D^{\dag }(\alpha )$. Second, let the system evolve according to
Eq.~\eqref{eq:U} and third, measure the polarization of the charge qubit.
Then
\begin{eqnarray*}
\langle \sigma _{z}\rangle &=&\mathrm{{Tr}_{x+CQ}\{U\,D^{\dag }(\alpha )\rho
(0)D(\alpha )U^{\dag }\sigma _{z}\}} \\
&=&\mathrm{{Tr}_{x+CQ}\{D^{\dag }(\alpha )\rho (0)D(\alpha )U^{\dag }\sigma
_{z}U\}} \\
&=&\mathrm{{Tr_{x}}\{D^{\dag }(\alpha )\rho_x D(\alpha )e^{i\pi a^{\dag
}a}\}=\frac{\pi}{2}W(\alpha )\,.}
\end{eqnarray*}
Due to the special properties of the evolution operator $U$, the summation of Eq.~\eqref{eq:Wigner1}
is performed by the system itself and therefore
only one value of $\langle\sigma_z\rangle$ already determines $W(\alpha)$.

The crucial point for the implementation of this method in mesoscopic systems
is the construction of the time evolution $%
U $ as given in Eq.~\eqref{eq:U}. It can be achieved with a Hamiltonian of the form~\cite{Davidovich}
\begin{equation}
H=\frac{1}{2}(E_J+\hbar \Delta(t) a^\dag a)\sigma_z + \hbar \nu a^\dag a.
\end{equation}
Then a Ramsey interferometry produces the result of Eq.~\eqref{eq:U}
if the waiting time, $\tau$ satisfies $\Delta\tau=\pi$. In Ref.~\cite{Schwab_Measure} it has been
pointed out that a static coupling between the resonator and the CPB
leads to the desired shift of the energy splitting of the charge qubit,
\begin{equation}
E_{e,n}-E_{g ,n}=E_{J}-2|\lambda |^{2}\frac{(2n+1)E_{J}}{%
E_{J}^{2}-(\hbar \nu )^{2}}\,.
\end{equation}%
The number state dependent part of the energy difference, $n\Delta
=n\,4|\lambda |^{2}E_{J}/(E_{J}^{2}-(\hbar \nu )^{2})$, can therefore be used to
implement $U$. For the parameter values given in the previous parts of this paper
($\lambda \approx 10$ MHz, $\nu\approx 100$ MHz, $E_{J}\approx 10$ GHz) the shift $\Delta
=40 $ kHz is actually too small but by optimizing the system parameters values for
$\Delta$ up to $4$ MHz~\cite{Schwab_Measure} are possible.
With such a system the evolution, U can be performed within the decoherence time, $T_2$
of the charge qubit~\cite{Vion} which would allow the implementation of this fast tomography method.

\end{appendix}

%%%%%%%%%%%%%%%%%%%%%%%%%%%%%%%%%%%%%%%%%%%%%%%%%%%%%%%%%%%%%%%%%%%%%%%%%%%%

%                                   Bib

%%%%%%%%%%%%%%%%%%%%%%%%%%%%%%%%%%%%%%%%%%%%%%%%%%%%%%%%%%%%%%%%%%%%%%%%%%%%%

\bibliographystyle{apsrev}
\bibliography{ref}

\end{document}